\documentclass[twocolumn,superscriptaddress,preprintnumbers,amsmath,amssymb,floatfix]{revtex4}

\usepackage{graphicx}
\usepackage{color}

\begin{document}

\title{Quasiclassical nonlinear plasmon resonance in graphene}

\author{Marinko Jablan}
\email{mjablan@phy.hr}
\affiliation{Department of Physics, Faculty of Science, University of Zagreb, 10000 Zagreb, Croatia}

\date{\today}

\begin{abstract}
Electrons in graphene behave like relativistic Dirac particles which can reduce velocity of light by two orders of magnitude in the form of plasmon-polaritons. Here we show how these properties lead to a peculiar nonlinear plasmon response in the quasiclassical regime of terahertz frequencies. On one hand we show how interband plasmon damping is suppressed by the relativistic Klein tunneling effect. On the other hand we demonstrate huge enhancement of the nonlinear intraband response when plasmon velocity approaches the resonance with the electron Fermi velocity. This extreme sensitivity on the plasmon intensity could be used for new terahertz technologies. \end{abstract}

\maketitle


\section{Introduction}

Nonlinear optics holds promise for the development of ultrafast information processing devices, however it typically requires huge optical intensities \cite{Cotter1999}. There is a constant search for new materials with stronger nonlinear response so there was naturally a huge interest in the nonlinear properties of the recently discovered material graphene \cite{Novoselov2005,Mikhailov2008,Mishchenko2009,Bao2009,Aoki2009,Ishikawa2010,Glazov2011,Zhang2012,Gullans2013,Dignam2014,
Jablan2015,MacLean2015,Jadidi2016,Wang2016,Mikhailov2017,Hommelhoff2017,Ooi2017,Cox2017,
Pedersen2017,Corkum2017,Tanaka2017,Turchinovich2018,
Eliasson2018,Sun2018,Jian2019,Tollerton2019,Principi2019,Cox2019,Gonclaves2020}.
Particularly it was argued that graphene has a strong nonlinear response in the form of interband multiplasmon absorption \cite{Jablan2015}. However this is a perturbative process, which strictly speaking makes sense only if $N+1$ plasmon absorption is much less than $N$ plasmon absorption. 
In this paper we wish to discuss what happens at THz frequencies due to many exciting applications in spectroscopy, security and wireless communications \cite{Tonouchi2007}.
For such low frequencies, the perturbative approach of multiplasmon absorption breaks down since it gets increasingly harder to distinguish $N$ from $N+1$ plasmon absorption if $N\gg 1$. On the other hand, since then electric field changes extremely slowly in time, process can be better understood as the quasiclassical tunneling. 
Here we provide a general model that can describe graphene response to strong electromagnetic field, and solve the model explicitly in the quasiclassical case of slow oscillations in space and time. We are particularly interested in the plasmon-polariton modes which can reduce velocity of light by two orders of magnitude \cite{Jablan2009}. We show a huge enhancement of nonlinear intraband response when plasmon velocity approaches the resonance with the electron Fermi velocity, while interband plasmon damping is surprisingly suppressed by the Klein tunneling effect \cite{Katsnelson2006}. Both effects crucially depend on the massles Dirac Hamiltonian so we first have to solve the Dirac equation in a strong electromagnetic field. In section \ref{2} we discuss the Quasiclassical approximation and in section \ref{3} we discuss interband dynamics beyond the Quasiclassical approximation. In section \ref{4} we calculate the general nonlinear current response and use this result to analyze interband dissipation in section \ref{5} and intraband resonance in section \ref{6}. Finally in section \ref{7} we provide discussion and conclusion. 

\section{Quasiclassical Dirac states in a strong electromagnetic field}
\label{2}

Electron motion in graphene is governed by a Dirac Hamiltonian:
\begin{equation}
{\hat{H}=v_F\mbox{\boldmath $\sigma$} \cdot {\bf {\hat p}}},
\label{dirac_hamil}
\end{equation}
where $v_F=10^6$~m/s is the Fermi velocity, ${\bf {\hat p}} =-i\hbar\mbox{\boldmath $\nabla$}$ is momentum operator, $\mbox{\boldmath $\sigma$}=(\sigma_x,\sigma_y)$, and $\sigma_{x,y}$ are Pauli spin matrices \cite{CastroNeto2009}. The corresponding eigenstates are:
\begin{equation}
\hat{H}\Psi^0_{{\bf P}_n}=E_{{\bf P}_n}\Psi^0_{{\bf P}_n},
\label{eigenstate}
\end{equation}
\begin{equation}
\Psi^0_{{\bf P}_n}({\bf r},t)= 
\frac{1}{\sqrt{2L^2}}
\left( {\begin{array}{c}
e^{-\frac{i}{2} \Phi_{{\bf P}_n}} \\
ne^{\frac{i}{2} \Phi_{{\bf P}_n}} \\
\end{array}} \right)
e^{\frac{i}{\hbar} ({\bf P}_n\cdot{\bf r}-E_{{\bf P}_n}t)}.
\label{freestate}
\end{equation}
Here ${\bf r}=(x,y)$, $L^2$ is the area of graphene flake, the electron momentum is ${\bf P}_n=(p_n,p_y)$, and the phase:
\begin{equation}
e^{i\Phi_{{\bf P}_n}}=\frac{p_n+ip_y}{|{\bf P}_n|}.
\label{freephase}
\end{equation}
Note that electron energies (eigenvalues) show a peculiar linear dispersion:
\begin{equation}
E_{{\bf P}_n}=nv_F|{\bf P}_n|,  
\label{freeenergy}
\end{equation}
where $n=-1$ represents the valence band, and $n=1$ the conduction band.  

To describe behavior of graphene in external vector potential ${\bf A}({\bf r},t)$ we need to solve the Dirac equation: 
\begin{equation}
i\hbar\frac{\partial\Psi}{\partial t}=v_F\mbox{\boldmath $\sigma$} \cdot ({\bf {\hat p}}-e{\bf A})\Psi. 
\label{dirac_eq}
\end{equation}
Particularly we are interested in longitudinal field: 
\begin{equation}
{\bf A}({\bf r},t)={\bf e}_x A({\bf r},t)={\bf e}_x A_0 \sin{u}, 
\label{field_harmonic}
\end{equation}
where $u=\omega t-qx$, and ${\bf e}_x$ is unit vector in the x direction. This field can then describe plasmon-polariton modes whose velocity is much smaller than the speed of light $\omega/q\ll c$ \cite{Jablan2009} . The case of Dirac particles in the transverse field at the light line $\omega=qc$ was solved by Volkov \cite{Volkov1935}, but unfortunately this approach does not work in our case. On the other hand, since we are primarily interested in slow oscillations in space and time, we can search for a solution in the form of the quasiclassical state: 
\begin{equation}
\Psi^{qc}=a e^{\frac{i}{\hbar}S}, 
\label{qc}
\end{equation}
where $S$ is the classical action and $a$ is the slowly varying amplitude \cite{LLQM}. Moreover we will see that these states enable us to get a much more general description of the system, including the fast oscillations in space and time. As a lowest approximation, let us insert this ansatz into Dirac equation and neglect terms containing $\hbar$, which is an excellent approximation in the case of slow oscillations, i.e. for $\hbar\omega\ll E_F$, and $\hbar q\ll p_F$, where $E_F$ is the Fermi energy and $p_F=E_F/v_F$ is the Fermi momentum. We obtain the equation of motion: 
\begin{equation}
-\frac{\partial S}{\partial t}a=v_F\mbox{\boldmath $\sigma$} \cdot (\mbox{\boldmath $\nabla$}S-e{\bf A})a, 
\label{qc_eq_motion}
\end{equation}
which is solved by the following quasiclassical states: 
\begin{equation}
\Psi_{{\bf P}_n}^{qc}({\bf r},t)= 
\frac{1}{\sqrt{2L^2}}
\left( {\begin{array}{c}
e^{-\frac{i}{2} \Phi_{{\bf P}_n^c-e{\bf A}}} \\
ne^{\frac{i}{2} \Phi_{{\bf P}_n^c-e{\bf A}}} \\
\end{array}} \right)
e^{\frac{i}{\hbar}S_{{\bf P}_n}},
\label{qcstate}
\end{equation}
where $S_{{\bf P}_n}$ satisfies the Hamilton-Jacobi equation for the classical action of the Dirac particle: 
\begin{equation}
\frac{\partial S_{{\bf P}_n}}{\partial t}=-nv_F\left|\mbox{\boldmath $\nabla$}S_{{\bf P}_n}-e{\bf A}\right|, 
\label{HamiltonJacobi}
\end{equation}
and we have introduced the classical momentum ${\bf P}_n^c=\mbox{\boldmath $\nabla$}S_{{\bf P}_n}$ \cite{LLM}. Since $y$ is a cyclic variable, the momentum is conserved in the $y$-direction and we can write ${\bf P}_n^c=(p_n^c,p_y)$, where $p_n^c=\partial S_{{\bf P}_n}/\partial x$. The phase is defined as: 
\begin{equation} 
e^{i\Phi_{{\bf P}_n^c-e{\bf A}}}=\frac{p_n^c-eA+ip_y}{|{\bf P}_n^c-e{\bf A}|}. 
\label{qcphase}
\end{equation}
We assume that the field is slowly turned on: 
\begin{equation}
A({\bf r},t)= A_0 \sin{(\omega t-qx)}e^{\eta t}, 
\label{adiabatic}
\end{equation}
where $\eta\ll\omega$, so that our quasiclassical state (\ref{qcstate}) adiabatically evolves from the free particle state (\ref{freestate}), i.e. we set the initial condition to be: 
\begin{equation}
{\Psi_{{\bf P}_n}^{qc}({\bf r},t=-\infty)}=\Psi^0_{{\bf P}_n}({\bf r},t). 
\label{initialcondition}
\end{equation}
To solve the Hamilton-Jacobi equation we use the ansatz \cite{LLCTF}: 
\begin{equation}
S_{{\bf P}_n}({\bf r},t)={\bf P}_n\cdot{\bf r}-E_{{\bf P}_n} t+F_{{\bf P}_n}(u),
\label{action_ansatz}
\end{equation}
which gives the following equation for the unknown function $\dot{F}=\frac{dF}{du}$: 
\begin{equation}
-E_{{\bf P}_n}+\omega\dot{F}_{{\bf P}_n}=-nv_F\sqrt{(p_n-q\dot{F}_{{\bf P}_n}-eA)^2+p_y^2}. 
\label{F}
\end{equation}
It is simple to solve this quadratic equation and obtain the classical momentum and energy:
\begin{equation}
p_n^c=\frac{\partial S_{{\bf P}_n}}{\partial x}=p_n-q\dot{F}_{{\bf P}_n}
\label{moment_def}
\end{equation}
\begin{equation}
E_{{\bf P}_n}^c=-\frac{\partial S_{{\bf P}_n}}{\partial t}=E_{{\bf P}_n}-\omega\dot{F}_{{\bf P}_n}
\label{energ_def}
\end{equation}
explicitly as:
\begin{equation}
\begin{split}
&p_n^c-eA=\frac{1}{1-\frac{q^2v_F^2}{\omega^2}}\left( p_n-\frac{q}{\omega}E_{{\bf P}_n}-eA\right.+\\
&n\frac{qv_F}{\omega}\left.\sqrt{\left(p_n-\frac{q}{\omega}E_{{\bf P}_n}-eA\right)^2+p_y^2\left(1-\frac{q^2v_F^2}{\omega^2}\right)}\right),
\label{moment}
\end{split}
\end{equation}
and:
\begin{equation}
\begin{split}
E_{{\bf P}_n}^c=\frac{v_F}{1-\frac{q^2v_F^2}{\omega^2}}\left( \frac{qv_F}{\omega}\left(p_n-\frac{q}{\omega}E_{{\bf P}_n}-eA\right)\right.+\\
\left.n\sqrt{\left(p_n-\frac{q}{\omega}E_{{\bf P}_n}-eA\right)^2+p_y^2\left(1-\frac{q^2v_F^2}{\omega^2}\right)}\right).
\label{energy}
\end{split}
\end{equation}
To check that initially: $p_n^c=p_n$, and $E_{{\bf P}_n}^c=E_{{\bf P}_n}$, one can note that $A({\bf r},t=-\infty)=0$ and use the following identity: 
\begin{equation}
\sqrt{\left(p_n-\frac{q}{\omega}E_{{\bf P}_n}\right)^2+p_y^2\left(1-\frac{q^2v_F^2}{\omega^2}\right)}=n\left( \frac{E_{{\bf P}_n}}{v_F} -\frac{qv_F}{\omega}p_n\right). 
\label{identity}
\end{equation}
Finally by using equations (\ref{action_ansatz}) and (\ref{energ_def}) it is convenient to write the action implicitly as:
\begin{equation}
{S_{{\bf P}_n}=\left(p_n-\frac{q}{\omega}E_{{\bf P}_n}\right) x+p_y\cdot y-\frac{1}{\omega}\int_0^u E_{{\bf P}_n}^c du}.
\label{action_implicit}
\end{equation}

\section{Interband dynamics beyond the quasiclassical approximation}
\label{3}

We can however get a much more general description of the system using these quasiclassical states (\ref{qcstate}). Let us start with some general wave-packet of the form 
\begin{equation}
\Psi({\bf r},t)=\sum_{n{{\bf P}_n}}c_{{\bf P}_n}(u)\Psi_{{\bf P}_n}^{qc}({\bf r},t), 
\label{wavepacket}
\end{equation}
and insert it into Dirac equation. It is then most convenient to consider the triplet $\{x,y,u\}$ as independent variables since $\{x,y\}$ variables appear only in the exponent $e^{\frac{i}{\hbar}S_{{\bf P}_n}}$. From Eq. (\ref{action_implicit}) we then see that our system dynamics can only couple states ${\bf P}_n$ and ${\bf P}_n'$ if: $p_y'=p_y$ and $p_n'-\frac{q}{\omega}E_{{\bf P}_n'}=p_n-\frac{q}{\omega}E_{{\bf P}_n}$. First condition is just the conservation of momentum in the $y$-direction, while the second condition corresponds to the multiphoton absorption process \cite{Jablan2015} which is given by the conservation of momentum: $p_n'-p_n=N\hbar q$, and conservation of energy: $E_{{\bf P}_n'}-E_{{\bf P}_n}=N\hbar \omega$. In this paper we consider only the case $qv_F/\omega<1$ since otherwise the (intraband) single-photon absorption dominates the system response \cite{Jablan2009,Jablan2015}. In this case it is straight forward to show that multiphoton absorption can couple only states in different bands $n'=-n$, i.e. second condition gives: 
\begin{equation}
p_{-n}-\frac{q}{\omega}E_{{\bf P}_{-n}}=p_n-\frac{q}{\omega}E_{{\bf P}_n}, 
\label{n-n}
\end{equation}
which can be solved as: 
\begin{equation}
p_{-n}=\frac{p_n\left(1+\frac{q^2v_F^2}{\omega^2}\right)-2\frac{q}{\omega}E_{{\bf P}_n}}{1-\frac{q^2v_F^2}{\omega^2}}. 
\label{p-n}
\end{equation}
It is also convenient to calculate the density of states: 
\begin{equation}
\frac{dp_{-n}}{dp_n}=-\frac{E_{{\bf P}_{-n}}}{E_{{\bf P}_n}}. 
\label{DOS}
\end{equation}
We see now that within our general wave-packet, states $\sum_n c_{{\bf P}_n}\Psi_{{\bf P}_n}^{qc}$ evolve completely independently from one another. Let us then focus on the state: 
\begin{equation}
\Psi_{{\bf P}_m}({\bf r},t)=\sum_{n=\pm m} c_{{\bf P}_n}(u)\Psi_{{\bf P}_n}^{qc}({\bf r},t),
\label{state_m_simple}
\end{equation}
subject to the initial condition: 
\begin{equation}
\Psi_{{\bf P}_m}({\bf r},t=-\infty)=\Psi^0_{{\bf P}_m}({\bf r},t),
\label{initialcondition_m}
\end{equation}
i.e. 
\begin{equation}
 c_{{\bf P}_m}(u=-\infty)=1, 
\label{initialcondition_cm}
\end{equation}
\begin{equation}
{c_{{\bf P}_{-m}}(u=-\infty)=0}. 
\label{initialcondition_c-m}
\end{equation}
We can further simplify calculations by writing the state in a more general form: 
\begin{equation}
\Psi_{{\bf P}_m}({\bf r},t)=\sum_{n=\pm m} c_{{\bf P}_n}(u)b_{{\bf P}_n}(u)\Psi_{{\bf P}_n}^{qc}({\bf r},t), 
\label{state_m}
\end{equation}
where we have introduced additional functions $b_{{\bf P}_n}(u)$ subject to initial condition: 
\begin{equation}
b_{{\bf P}_m}(u=-\infty)=1. 
\label{initialcondition_bm}
\end{equation}
Particularly by choosing: 
\begin{equation}
b_{{\bf P}_n}(u)=B_{{\bf P}_n}\sqrt{E_{{\bf P}_n}^c/\Delta E_{{\bf P}_n}^c}, 
\label{bn}
\end{equation}
where $\Delta E_{{\bf P}_n}^c=E_{{\bf P}_n}^c-E_{{\bf P}_{-n}}^c$, and 
\begin{equation}
B_{{\bf P}_m}=B_{{\bf P}_{-m}}=\sqrt{\Delta E_{{\bf P}_m}/E_{{\bf P}_m}}
\label{Bm}
\end{equation}
we obtain (see Appendix \ref{beyondQC}): 
\begin{equation}
|c_{{\bf P}_m}(u)|^2+|c_{{\bf P}_{-m}}(u)|^2=1.
\label{conservation_of_number}
\end{equation}
We can then interpret $|c_{{\bf P}_n}(u)|^2$ as the probability of finding the electron in the band $n$, as a function of $u$. However one needs to be careful about this interpretation since $u=\omega t-qx$, so this is not the standard probability as a function of time $t$. Finally, in the case of slow oscillations in space and time we can use Landau-Zener model \cite{LLQM} to obtain explicitly:
\begin{equation}
|c_{-m}(u=\infty)|^2=\exp\left({\frac{1}{\hbar\omega} \textrm{im} \int_{C} \Delta E_m^c du}\right)=K
\label{LandauZener},
\end{equation}
where the integration contour $C$ goes around the complex transition point $u_0$ which is given by $\Delta E_m^c(u_0)=0$. Here $K$ is the transition probability for a single passage while the probability for a double passage is $2K(1-K)$ \cite{LLQM} (see also Appendix \ref{LZ}).

\section{Nonlinear current response}
\label{4}

To describe the general case of mixed state we can introduce the density matrix: 
\begin{equation}
\rho({\bf r},t,{\bf r}',t')=4\sum_{n{\bf P}_n}f_{{\bf P}_n}\Psi^*_{{\bf P}_n}({\bf r}',t')\Psi_{{\bf P}_n}({\bf r},t), 
\label{DM}
\end{equation}
where $f_{{\bf P}_n}=\frac{1}{e^{(E_{{\bf P}_n}-E_F)/kT}+1}$ is the Fermi-Dirac distribution at temperature $T$ \cite{LLSP}, and  we took into account 2 spin and 2 valley degeneracy in graphene \cite{CastroNeto2009}. We can then write the induced current as \cite{LLQM}: 
\begin{equation}
{\bf j}({\bf r},t)=\int d{\bf R} \left[ \hat{\bf j}({\bf r})\rho({\bf R},t,{\bf R}',t) \right]_{{\bf R}'={\bf R}}, 
\label{current_mixed_state}
\end{equation}
where $\hat{\bf j}({\bf r})=ev_F\mbox{\boldmath $\sigma$}\delta(\hat{\bf r}-{\bf r})$ is the current density operator of graphene \cite{CastroNeto2009}. Since $j_y=0$ due to symmetry we can focus only on $x$-component:
\begin{equation} 
\begin{split}
j_x({\bf r},t)=\frac{4}{L^2}\sum_{n{\bf P}_n} f_{{\bf P}_n}&\left( |c_{{\bf P}_n}|^2 \, b_{{\bf P}_n}^2 e v_{{\bf P}_n}^c + \right. \\
+|&c_{{\bf P}_n{\bf P}_{-n}}|^2 \, b_{{\bf P}_n{\bf P}_{-n}}^2 e v_{{\bf P}_{-n}}^c+\\
+\frac{\Delta E_{{\bf P}_n}^c}{\dot A}\frac{B_{{\bf P}_n}^2}{2}&\left.\left( 1-\frac{q^2v_F^2}{\omega^2}\right) \frac{d|c_{{\bf P}_n{\bf P}_{-n}}|^2}{du}\right),
\end{split}
\label{current}
\end{equation}
where $\dot{A}=\frac{dA}{du}$, and $v_{{\bf P}_n}^c=\frac{\partial E_{{\bf P}_n}^c}{\partial p_n^c}=nv_F\cos{\Phi_{{\bf P}_n^c-e{\bf A}}}$ is the $x$-component of the classical velocity (see Appendix \ref{nonlinear}). We can consider that the state with initial condition $c_{{\bf P}_n}=1$ and $c_{{\bf P}_n{\bf P}_{-n}}=0$ evolves independently from the state with initial condition $c_{{\bf P}_{-n}}=1$ and $c_{{\bf P}_{-n}{\bf P}_n}=0$, after averaging over thermally randomized initial phases. 
We can now interpret first part of Eq. (\ref{current}) ($|c_n|^2b_n^2 ev_n^c$) as the current of the electrons that have stayed in their original band, second part as the current of the electrons that have jumped into different band, while the third part describes the actual interband transition process i.e. the energy dissipation. Note that we could choose $b_{{\bf P}_n}=1$ but in that case it is no longer true that $|c_{{\bf P}_n}|^2+|c_{{\bf P}_n{\bf P}_{-n}}|^2=1$, and interband part becomes much more complicated.  

\section{Interband dissipated power}
\label{5}

While Eq. (\ref{current}) is exact, it requires numerical evaluation of coefficients $c_{{\bf P}_n}(u)$ (see Eq. (\ref{cs})). However in the case of slow oscillations in space and time we can use the Landau-Zener model (\ref{LandauZener}). Let us first find the dissipated power $P=\int d{\bf r}\,{\bf j}\cdot{\bf E}=\int d{\bf r}\, j_x E_x$, where $E_x=-\frac{\partial A_x}{\partial t}=-\omega\dot{A}$. Since $\dot{A}=A_0 \cos{u}$, only the third interband part contributes to the dissipation:  
\begin{equation}
\begin{split}
P=4\sum_{{\bf P}_1} & (f_{{\bf P}_{-1}}-f_{{\bf P}_1})\frac{[\Delta E_{{\bf P}_1}^c(u)]_{min}}{\textrm{T}} \times \\
& 2K(1-K) \frac{B_{{\bf P}_1}^2}{2} \left( 1-\frac{q^2v_F^2}{\omega^2}\right)
\label{power}.
\end{split}
\end{equation}
Here we have used the following relation: 
\begin{equation}
\frac{dp_{-n}}{dp_{n}}\frac{B_{{\bf P}_{-n}}^2}{B_{{\bf P}_n}^2}=-\frac{dp_{-n}}{dp_n}\frac{E_{{\bf P}_n}}{E_{{\bf P}_{-n}}}=1,
\label{DOS2}
\end{equation}
which is a direct consequence of Eq. (\ref{DOS}). Also we used:
\begin{equation}
\frac{d|c_{{\bf P}_n{\bf P}_{-n}}|^2}{du}=2K(1-K)\delta(u-\xi),
\label{tunneling_transition}
\end{equation}
i.e. we assumed that transition happens at a real point $\xi$ when the gap is minimal: $\Delta E_{{\bf P}_1}^c(\xi)=[\Delta E_{{\bf P}_1}^c(u)]_{min}$ since then tunneling probability is largest (see also Appendix \ref{LZ}). 
We can now clearly see physical interpretation of every part of Eq. (\ref{power}): $f_{-1}-f_1$ is the Pauli principle, $[\Delta E_1^c]_{min}$ is the dissipated energy per oscillation period $\textrm{T}=2\pi/\omega$, and $2K(1-K) \frac{{B_1}^2}{2} \left( 1-\frac{q^2v_F^2}{\omega^2}\right)$ is the transition probability. As we noted $|c_n(u)|^2$ is not the actual probability at time $t$ since $u=\omega t-qx$. Only in the case of homogenous field: $q=0$, do we get that $K$ (i.e. $2K(1-K)$) is the transition probability in time for a single passage (i.e. double passage).
Note that $K$ from Eq. (\ref{LandauZener}) exponentially decreases as we increase the gap $[\Delta E_1^c]_{min}$. 
The leading contribution to the dissipated power then comes from the states near the lowest gap (minimum of $[\Delta E_1^c]_{min}$) i.e. for $p_y=0$ and $p_1=p_F$ since the Pauli principle requires that $p_1^2+p_y^2\geq p_F^2$ for $kT\ll E_F$. 
In that case: 
\begin{equation}
[\Delta E_1^c]_{min}=\frac{2v_F\left|p_F\left(1-\frac{qv_F}{\omega}\right)-eA_0\right|}{1-\frac{q^2v_F^2}{\omega^2}}, 
\label{mingap}
\end{equation}
and we see that at the threshold $A_0=A_{nl}$, where:
\begin{equation}
A_{nl}=\left(1-\frac{qv_F}{\omega}\right)\frac{p_F}{e}, 
\label{threshold}
\end{equation}
the gap disappears $[\Delta E_1^c]_{min}=0$, and we get a perfect tunneling $K=1$ for the single passage (just like the Klein tunneling effect \cite{Katsnelson2006}). However the particle simply returns back to the original band upon the return passage since $2K(1-K)=0$. In other words we expect to see that $P$ grows exponentially with $A_0$ until the threshold $A_{nl}$ when it starts to saturate. Finally since Klein tunneling does not result in energy dissipation ($[\Delta E_1^c]_{min}=0$), we get very small values for the total dissipated power (see figure \ref{fig1}(b)). We could also calculate dissipated power by a Keldysh approach \cite{Keldysh1965} however one has to specially deal with the close spaced singularities at the onset of Klein tunneling. 

\section{Intraband nonlinear resonance}
\label{6}

With the forementioned analysis in mind we can find the dominant contribution to the current (\ref{current}) by writing $|c_{{\bf P}_n}|\approx 1$, and $|c_{{\bf P}_n{\bf P}_{-n}}|\approx 0$, so that $j_x({\bf r},t)=\frac{4}{L^2}\sum_{n{\bf P}_n}f_{{\bf P}_n} \, b_{{\bf P}_n}^2 e v_{{\bf P}_n}^c$. If we then assume that $kT\ll E_F$ so that the valence band is completely occupied (and thus can not conduct electricity) we are left with the conduction band ($n=1$) current which can be written as (see Appendix \ref{nonlinear}): 
\begin{equation}
j_x=\frac{4e}{h^2}\int dp_1dp_y\,f_1 \frac{\partial E_1^c}{\partial p_1}=-\frac{4e}{h^2}\int dp_1dp_y \frac{\partial f_1}{\partial p_1}E_1^c. 
\label{intraband}
\end{equation}
Current (\ref{intraband}) is plotted in figure \ref{fig1}(c) for the local case $qv_F/\omega\approx 0$, and in figure \ref{fig1}(d) for the nonlocal case $qv_F/\omega\approx 1$. In the local case it is easy to visualize the result since the field uniformly shifts all electrons in momentum space: $p_1\rightarrow p_1-eA_0\sin{\omega t}$ (see the inset in figure \ref{fig1}(c)). Then due to peculiar linear Dirac dispersion, at the peak field for $eA_0\gg p_F$ majority of electrons reach the maximum electron velocity $v_F$ in graphene and the current saturates. While some of these intraband effects were discussed for the local case \cite{Mikhailov2008,Mikhailov2017}, we show a dramatic new physics in the nonlocal response. Particularly for $qv_F/\omega\approx 1$ current becomes extremely nonlinear since classical energy (\ref{energy}) is very asymmetric depending on the sign of the field: $E_1^c\propto \Theta\left( p_1-\sqrt{p_1^2+p_y^2}-eA\right)$. Particularly for $eA>0$ very little current flows compared to the $eA<0$ case, and our system behaves like a rectifier (see figure \ref{fig1}(d)). To reach this nonlinear response requires only that: $\left(p_1-\frac{q}{\omega}E_{{\bf P}_1}-eA\right)^2\gg p_y^2\left(1-\frac{q^2v_F^2}{\omega^2}\right)$. For $qv_F/\omega\approx 1$ this will be satisfied practically always if $eA_0\gg eA_{nl}=(1-qv_F/\omega)p_F$ (see Appendix \ref{linear} for the linear response regime $eA_0\ll eA_{nl}$).
Figures 1(c) and 1(d) show the case of the photon energy $\hbar\omega\approx E_F/3$, which for an electron concentration $n=\frac{p_F^2}{\pi\hbar^2}=10^{12}\,\textrm{cm}^{-2}$ corresponds to the frequency $\frac{\omega}{2\pi}\approx 9 \,\textrm{THz}$. At room temperature: $kT\approx0.2\, E_F$, so we can neglect temperature effects. Note that the threshold for the onset of nonlinear behavior: $A_{nl}=(1-qv_F/\omega)p_F/e$, decreases linearly with Fermi energy like in the local case \cite{Mikhailov2008,Mikhailov2017}. But what is especially intriguing is that $A_{nl}$ goes to zero at the resonance of plasmon velocity and the electron Fermi velocity in graphene, dramatically enhancing nonlinear response in the nonlocal case. This extreme sensitivity on the electric field amplitude and the rectifying effect shown in the figure \ref{fig1}(d) could be used for the detection of THz radiation, and in the more advanced applications, for information processing devices \cite{Cotter1999}. Of course, like in atomic resonances, the final scale of nonlinearity will be determined by the loss mechanisms. Graphene room temperature DC mobility can be larger than $\mu=10$ m$^2/$Vs \cite{Sarma2008,Bolotin2008} which corresponds to damping rate $\gamma=\frac{ev_F^2}{\mu E_F}\approx 0.9$ THz (see Appendix \ref{linear}). Since $\hbar\omega \sim kT$ this will not be drastically changed at THz frequencies. System response is then undetermined within the linewidth $\gamma/\omega\approx 0.01$ and so for $1-qv_F/\omega<0.01$ this theory has to be supplemented by taking losses into account.

\section{Discussion and conclusion}
\label{7}

\begin{figure}
\centerline{
\mbox{\includegraphics[width=0.5\textwidth]{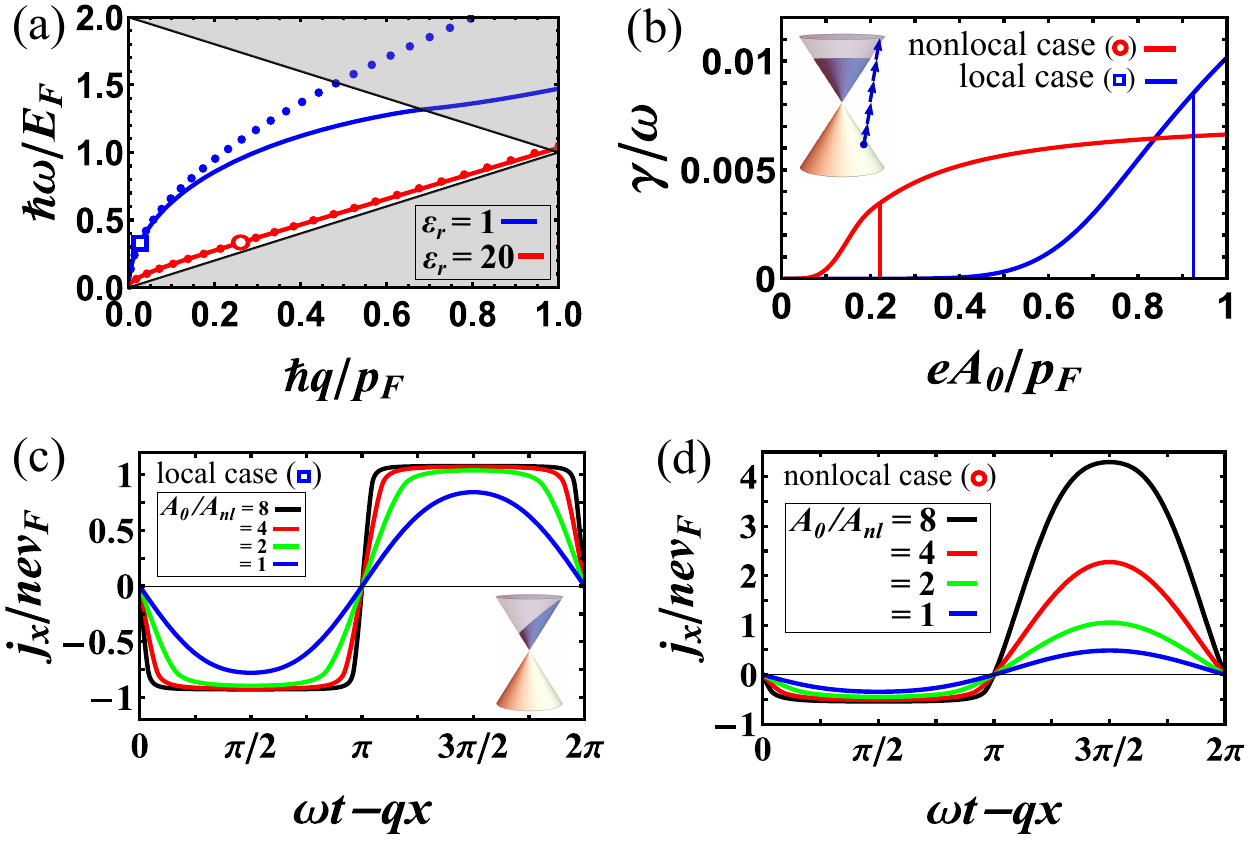}}
}
\caption{
(a) Plasmon dispersion in graphene for different dielectric environments $\varepsilon_{r}$. Solid lines: Random phase approximation \cite{Jablan2009}. Dots: quasiclassical linear response from Eq. (\ref{dispersion}). Gray area: regime of a single plasmon absorption i.e linear Landau damping. Open square represents a point for which $qv_F/\omega\approx0.08$ and local theory is applicable. Nonlinear response for this point is shown in figures (b) and (c). Open circle represents nonlocal case $qv_F/\omega\approx0.8$ for which nonlinear response is shown in figures (b) and (d).
(b) Quasiclassical nonlinear Landau damping: dependence of plasmon linewidth on the amplitude of the vector potential $A_0$. Vertical line represents the amplitude $A_0=A_{nl}=(1-q v_F/\omega)p_F/e$ and the onset of the Klein tunneling. Inset shows the multiplasmon absorption process, which for low frequencies $\hbar\omega\ll2E_F$, is better described as a quasiclassical Landau-Zener tunneling. (c) Intraband current response at different amplitudes $A_0$ for the local case. Inset shows snapshot of the electron dynamics. (d) Intraband nonlocal current response.
}
\label{fig1}
\end{figure}

For small fields $eA_0\ll eA_{nl}$ we can linearize the current (\ref{intraband}) to obtain: $j_x=i\omega\sigma(q,\omega)A$.
However oscillating current will also induce vector potential that will act back on the current. If we then introduce some external potential $A^{ext}$, the current will respond not only to $A^{ext}$ but to the total self-consistent potential $A$ of the amplitude $A_0=\frac{A_0^{ext}}{1+\frac{iq\sigma(q,\omega)}{\omega 2\varepsilon_0\varepsilon_r}}$ (see Appendix \ref{linear}). One can see that it is possible to have self-sustained oscillations of the electron gas (plasmon-polaritons) even in the absence of the external field if: $1+\frac{iq\sigma(q,\omega)}{\omega 2\varepsilon_0\varepsilon_r}=0$, with the corresponding plasmon dispersion $\omega(q)$ plotted in figure \ref{fig1}(a). Furthermore we see that we get huge enhancement of the external field at the plasmon resonance. Note that this analysis gets much more complicated for large fields $eA_0\gg eA_{nl}$ since the current response is extremely nonlinear and it will produce vector potential with many new harmonics (see figures \ref{fig1}(c) and \ref{fig1}(d)), while our calculation is based on the single harmonic in the vector potential $A=A_0\sin{(\omega t - qx)}$. The precise analysis including the full self-consistent nonlinear effects simply goes beyond the scope of this paper and here we can only discuss some qualitative properties. 
Generation of direct current or higher harmonics, which all extract energy from the basic harmonic, would be manifested as effective plasmon damping. Therefore one would see increase in the plasmon linewidth due to pure intraband effects in addition to interband dissipation process. There would also be an intensity dependent response at the basic harmonic which would shift the plasmon dispersion. Yet especially interesting case occurs for large $\varepsilon_r$ when both the plasmon dispersion and higher harmonics lie close to the line $\omega= qv_F$, since then self-consistent effects would additionally enhance higher harmonics. 
It is interesting to note that in that case all harmonics separately will show similar behavior with the similar threshold field $eA_{nl}=(1-qv_F/\omega)p_F$, but there might occur particularly strong nonlinear interaction between these harmonics. On the other hand, to test the quantitative predictions of this work for large fields $eA_0\gg eA_{nl}$ it would be most simple to measure the DC component of the current (\ref{intraband}) for nonlocal excitation $\omega\gtrsim qv_F$, making sure that none of the harmonics cuts the plasmon dispersion.

To quantify interband plasmon dissipation it is most simple to look at the dissipation rate: $\gamma=P/W$, where $W$ is the total plasmon energy. 
The energy density of a dispersive medium can be written as \cite{LLECM}: ${u=\frac{1}{2}\mbox{re}\frac{d(\omega\varepsilon)}{d\omega}\langle {\bf E}^2 \rangle}$, which in the case of graphene plasmons gives \cite{Jablan2015}: 
\begin{equation}
\frac{W}{L^2}=\frac{A_0^2\omega^3 }{4}\frac{d}{d\omega}\left( \frac{-\textrm{im}\sigma(q,\omega)}{\omega} \right). 
\label{plasmonenergy}
\end{equation}
One can then write the plasmon linewidth as $\gamma/\omega=P/\omega W$ which basically says what fraction of the plasmon energy is dissipated during a single oscillation period. While plasmon linewidth is very small it can be none the less detected in precise measurements due to very specific dependence on the plasmon amplitude $A_0$. For small amplitudes we see exponential growth with $A_0$ typical of the quasiclassical tunneling, while for large amplitudes we see saturation effect which signals the onset of Klein tunneling (figure \ref{fig1}(b)). We call this effect quasiclassical nonlinear Landau damping to distinguish it from the nonlinear  Landau damping discussed recently in classical plasma at high frequencies \cite{Villani2011}. Note that expression (\ref{power}) doesn't represent truly dissipated energy, but more like a stored energy that can be retrieved back from the system. One spectacular way in which this can happen is if after the electron has tunneled into a different band, it gets accelerated by this strong electric field and finally recombines with the hole it left behind, liberating this huge energy from the field in the form of a train of high harmonics \cite{Corkum1993,Lewenstein1994}. While this too is a very weak effect it shows intriguing properties in the frequency space. Namely this train of harmonics adds up to a pulse extremely localized in time on the order of atto seconds \cite{Lewenstein1996}. Effect that would be even more interesting with plasmons in graphene due to their subwavelenght nature \cite{Jablan2009} since the resulting pulse would be localized in time and space. While high harmonic generation with plasmons in graphene was analyzed numerically \cite{Cox2017}, our quasiclassical states offer the most natural platform to take into account the quasiclassical nature of this problem \cite{Lewenstein1994}.

In conclusion we have developed a general model that can treat the response of graphene to a strong electromagnetic field, which we explicitly solved in the quasiclassical regime of THz frequencies. Interband transitions are analyzed via the Landau-Zener model, leading to plasmon dissipation which is however suppressed by the Klein tunneling effect. 
Moreover our quasiclassical states could be further used to find how this dissipated energy can be extracted back via the three step process of high harmonic generation \cite{Corkum1993,Lewenstein1994,Lewenstein1996}. Most notably we demonstrate huge enhancement of nonlinear intraband response near the resonance of plasmon velocity and electron Fermi velocity in graphene. This extreme sensitivity on the plasmon intensity could be used for nonlinear, subwavelenght THz technologies like detectors or information processing devices. 

This work was supported by University of Zagreb (Research support no. 20283205), and QuantiXLie Centre of Excellence, a project cofinanced by the Croatian Government and European Union through the European Regional Development Fund - the Competitiveness and Cohesion Operational Programme (Grant KK.01.1.1.01.0004).

\appendix

\section{Interband dynamics beyond the quasiclassical approximation}
\label{beyondQC}

To simplify notation let us write our state ${\Psi_{{\bf P}_m}({\bf r},t)=\sum_n c_{{\bf P}_n}(u)b_{{\bf P}_n}(u)\Psi_{{\bf P}_n}^{qc}({\bf r},t)}$ as: 
\begin{equation}
{\Psi_m=\sum_{n=\pm m} c_n b_n a_n e^{\frac{i}{\hbar}S_n}}, 
\label{psi_m}
\end{equation}
where we have used that: ${\Psi^{qc}_n({\bf r},t)=a_n(u)e^{\frac{i}{\hbar}S_n}}$. We use this ansatz to solve the Dirac equation: $i\hbar\frac{\partial\Psi}{\partial t}={v_F\mbox{\boldmath $\sigma$} \cdot ({\bf {\hat p}}-e{\bf A})\Psi}$. 
By choosing $a_n,S_n$ to satisfy the quasiclassical equation of motion: $-\frac{\partial S_n}{\partial t}a_n={v_F\mbox{\boldmath $\sigma$} \cdot (\mbox{\boldmath $\nabla$}S_n-e{\bf A})a_n}$, we obtain the equation for the remaining unknowns: ${(\omega-qv_F\sigma_x)\sum_n\frac{d}{du}(c_n b_n a_n)e^{\frac{i}{\hbar}S_n}=0}$. Since the matrix: 
\begin{equation}
M=\omega-qv_F\sigma_x=\left( \begin{array}{cc} \omega & -qv_F \\ -qv_F & \omega \\  \end{array} \right),
\label{matrix}
\end{equation}
is invertible for $\omega\neq qv_F$, we multiply previous equation by $M^{-1}$ to obtain the (exact) evolution equation:
\begin{equation}
\sum_n\frac{d}{du}(c_n b_n a_n)e^{\frac{i}{\hbar}S_n}=0. 
\label{evolutioneq}
\end{equation}
We can enormously simplify further calculations by choosing the function $b_n(u)$ so that $\frac{d(b_n a_n)}{du}\propto b_{-n}a_{-n}$, which means that we maximally decouple dynamics between the bands. 
It is easy to solve this equation via the substitution $b_n=e^{\beta_n}$ to obtain: $b_n(u)=B_n\sqrt{E_n^c/\Delta E_n^c}$, and here we give a short check of the solution. 
Let us focus on a spinor: 
\begin{equation}
d_n=b_n a_n =
\frac{B_n}{\sqrt{2L^2}}\sqrt{\frac{E_n^c}{\Delta E_n^c}}
\left( {\begin{array}{c}
e^{-\frac{i}{2} \Phi_{{\bf P}_n^c-e{\bf A}}} \\
ne^{\frac{i}{2} \Phi_{{\bf P}_n^c-e{\bf A}}} \\
\end{array}} \right)
\label{d1}.
\end{equation}
Then since: $E_n^c=nv_F\sqrt{(p_n^c-eA)^2+p_y^2}$, we can write $e^{i\Phi_{{\bf P}_n^c-e{\bf A}}}=(p_n^c-eA+ip_y)nv_F/E_n^c$, and:
\begin{equation}
\begin{split}
d_n=&
\frac{B_n}{\sqrt{2L^2}}\sqrt{\frac{v_F}{n\Delta E_n^c}}
\left( {\begin{array}{c}
\sqrt{p_n^c-eA-ip_y} \\
n\sqrt{p_n^c-eA+ip_y} \\
\end{array}} \right) \\
=&
\frac{B_n}{\sqrt{2L^2}}
\left( {\begin{array}{c}
D_n \\
nD_n^* \\
\end{array}} \right).
\end{split}
\label{d2}
\end{equation}
Here we have used the fact that $nE_n^c$ and $n\Delta E_n^c$ are positive quantities, and we have introduced a function:
\begin{equation}
{D_n=
\sqrt{\frac{p_n-\frac{q}{\omega}E_n-eA-ip_y\left(1-\frac{q^2v_F^2}{\omega^2}\right)}{2\sqrt{\left(p_n-\frac{q}{\omega}E_n-eA\right)^2+p_y^2\left(1-\frac{q^2v_F^2}{\omega^2}\right)}}+n\frac{qv_F}{\omega}}}.
\label{D}
\end{equation}
If we then choose $B_{-m}=B_m=\mbox{constant}$, it is  straight forward to show that: 
\begin{equation}
\begin{split}
\dot{d_n}&=d_{-n}\frac{-\frac{i}{2}p_ye\dot{A}\sqrt{1-\frac{q^2v_F^2}{\omega^2}}}{\left(p_n-\frac{q}{\omega}E_n-eA\right)^2+p_y^2\left(1-\frac{q^2v_F^2}{\omega^2}\right)}\\
&=d_{-n}\frac{-2iv_F^2 p_y e\dot{A}}{(\Delta E_n^c)^2} \left( 1-\frac{q^2v_F^2}{\omega^2}\right)^{-3/2}.
\end{split}
\label{d3}
\end{equation}
If we now insert this expression into the evolution Eq. (\ref{evolutioneq}): $\sum_n\left( \dot{c_n} d_n+ c_n\dot{d_n}\right)e^{\frac{i}{\hbar}S_n}=0$, we obtain the following relations for the coefficients $c_n(u)$:
\begin{equation}
\dot{c}_{-n}=ic_n e^{\frac{i}{\hbar}(S_n-S_{-n})}\frac{2v_F^2 p_y e\dot{A}}{(\Delta E_n^c)^2}\left( 1-\frac{q^2v_F^2}{\omega^2} \right)^{-3/2}
\label{cs}.
\end{equation}
We can now immediately see that: $\frac{d|c_n|^2}{du}=-\frac{d|c_{-n}|^2}{du}$, and since initial conditions are set to: ${c_m(u=-\infty)=1}$ and $c_{-m}(u=-\infty)=0$, we obtain:
\begin{equation}
|c_m(u)|^2+|c_{-m}(u)|^2=1
\label{total}
\end{equation}

\section{Nonlinear current response}
\label{nonlinear}

Let us find the $\Psi_m$ contribution to the current density:
\begin{equation}
\begin{split}
j_x=ev_F\Psi_m^*\sigma_x\Psi_m &=ev_F\sum_{n=\pm m}\left(|c_n|^2d_n^*\sigma_x d_n+ \right. \\ 
&\left.+c_{-n}^*c_ne^{\frac{i}{\hbar}(S_n-S_{-n})}d_{-n}^*\sigma_xd_n\right).
\end{split}
\label{current1}
\end{equation}
From Eq. (\ref{d1}) we can write the intraband matrix element: 
\begin{equation}
d_n^*\sigma_x d_n = \frac{b_n^2}{L^2}n\cos\Phi_{{\bf P}_n^c-e{\bf A}}
=\frac{b_n^2}{L^2}\frac{v_n^c}{v_F} 
\label{intra_elem1}
\end{equation}
where we have introduced the classical velocity $v_n^c=\frac{\partial E_n^c}{\partial p_n^c}=nv_F\cos\Phi_{{\bf P}_n^c-e{\bf A}}$. Alternatively, using the equations (\ref{d2}) and (\ref{D})  we can express the same matrix element as:
\begin{equation}
\begin{split}
&d_n^*\sigma_x d_n = n\frac{B_n^2}{L^2}\textrm{re}(D_n^2)=n\frac{B_n^2}{2L^2}\times\\ 
&\left(\frac{p_n-\frac{q}{\omega}E_n-eA}{\sqrt{\left(p_n-\frac{q}{\omega}E_n-eA\right)^2+p_y^2\left(1-\frac{q^2v_F^2}{\omega^2}\right)}}
+n\frac{qv_F}{\omega}\right),
\label{intra_elem2}
\end{split}
\end{equation}

Next, using equations (\ref{d2}) and (\ref{D}) we can write the interband matrix element:
\begin{equation}
\begin{split}
d_{-n}^*\sigma_xd_n&=-n\frac{B_n^2}{L^2}\,i\,\textrm{im}(D_{-n}D_n)\\
&=\frac{B_n^2}{2L^2}\frac{2iv_F p_y}{\Delta E_n^c}\left( 1-\frac{q^2v_F^2}{\omega^2} \right)^{-1/2}
\end{split}
\label{inter_elem}
\end{equation} 
From equations (\ref{cs}) and (\ref{inter_elem}) we then obtain:
\begin{equation}
\begin{split}
c_{-n}^*c_ne^{\frac{i}{\hbar}(S_n-S_{-n})}d_{-n}^*\sigma_xd_n+c.c.=\\
=\frac{B_n^2}{2L^2}\frac{\Delta E_n^c}{v_Fe\dot{A}}\left(1-\frac{q^2v_F^2}{\omega^2}\right)\frac{d|c_{-n}|^2}{du}. 
\end{split}
\end{equation}
Finally we can write the current density (\ref{current1}) as:
\begin{equation}
\begin{split}
j_x=\frac{1}{L^2}\left(|c_m|^2b_m^2ev_m^c+|c_{-m}|^2b_{-m}^2ev_{-m}^c+ \right.\\ 
\left.\frac{\Delta E_m^c}{\dot{A}}\frac{B_m^2}{2}\left(1-\frac{q^2v_F^2}{\omega^2}\right)\frac{d|c_{-m}|^2}{du}\right)
\end{split}
\label{current2}
\end{equation}
In the quasiclassical case of low frequencies interband transitions are exponentially suppressed and we can approximately write: $c_m\approx1$, $c_{-m}\approx0$, so that the current density is: $j_x=b_m^2ev_m^c/L^2$. We note that this reduces to the classical single-band result $j_x^c=ev_m^c/L^2$ only in the nonlinear local case ($q=0$) or in linear nonlocal case. In other words $b_m$ amounts to quantum nonlinear, nonlocal, interband correction. By using density matrix it is straight forward to generalize this to the case of the electron Fermi see described by the Fermi-Dirac distribution $f_m=\frac{1}{e^{(E_m-E_F)/kT}+1}$ as: 
\begin{equation}
j_x=\frac{4}{L^2}\sum_{p_mp_y} f_mb_m^2ev_m^c =\frac{4}{h^2}\int dp_mdp_y \,f_mb_m^2ev_m^c
\label{current3}
\end{equation}
By using equations (\ref{intra_elem1}) and (\ref{intra_elem2}) we can write this in alternative form as:
\begin{equation}
\begin{split}
&j_x=\frac{4ev_F}{h^2}\int dp_mdp_y\,f_m \frac{B_m^2}{2}m \times \\
&\left( \frac{p_m-\frac{q}{\omega}E_m-eA}{\sqrt{\left(p_m-\frac{q}{\omega}E_m-eA\right)^2+p_y^2\left(1-\frac{q^2v_F^2}{\omega^2}\right)}} +m\frac{qv_F}{\omega} \right)
\end{split}
\label{current3}.
\end{equation}
Initial condition $b_m(u=-\infty)=1$ requires that: $B_m=\sqrt{\Delta E_m/E_m}$ where $\Delta E_m=E_m-E_{-m}$. From Eq. (\ref{n-n}) and (\ref{p-n}) it is straight forward to show that: 
\begin{equation}
\frac{B_m^2}{2}=\frac{1-m\frac{qv_F}{\omega}\frac{p_m}{\sqrt{p_m^2+p_y^2}}}{1-\frac{q^2v_F^2}{\omega^2}}
\label{B}.
\end{equation}
Eq. (\ref{current3}) can then be rewritten in a more convenient form:
\begin{equation}
j_x=\frac{4e}{h^2}\int dp_1dp_y\,f_1 \frac{\partial E_1^c}{\partial p_1}=
-\frac{4e}{h^2}\int dp_1dp_y \frac{\partial f_1}{\partial p_1}E_1^c
\label{current4},
\end{equation}
the last equation obtained by partial integration and we assumed that we are dealing with the conduction band $m=1$. Furthermore, in the low temperature case $kT\ll E_F$ we can write: $f_1=\Theta\left(p_F-\sqrt{p_1^2+p_y^2}\right)$, so that $-\frac{\partial f_1}{\partial p_1}=\sum_s s\delta(p_1-p_1^s)$, where $p_1^s=s\sqrt{p_F^2-p_y^2}$ and $s=\pm 1$. We can then evaluate one integral from Eq. (\ref{current4}) to obtain the current:
\begin{equation}
\begin{split}
j_x&=\frac{8ev_F}{h^2\left(1-\frac{q^2v_F^2}{\omega^2}\right)}\int_0^{p_F}dp_y 
\left( 2\frac{qv_F}{\omega}\sqrt{p_F^2-p_y^2}  + \right. \\  
&\sqrt{\left(\sqrt{p_F^2-p_y^2}-\frac{qv_F}{\omega}p_F-eA\right)^2+p_y^2\left(1-\frac{q^2v_F^2}{\omega^2}\right)}  \\ 
 - & \left. \sqrt{\left(-\sqrt{p_F^2-p_y^2}-\frac{qv_F}{\omega}p_F-eA\right)^2+p_y^2\left(1-\frac{q^2v_F^2}{\omega^2}\right)}\right)
\end{split}
\label{current5}.
\end{equation}

\section{Linear response regime}
\label{linear}

For small fields $eA_0\ll eA_{nl}=p_F(1-\frac{qv_F}{\omega})$ we can linearize the current (\ref{current5}) to obtain: $j_x=i\omega\sigma(q,\omega)A$, where the conductivity $\sigma(q,\omega)$ can be evaluated explicitly:
\begin{equation}
\sigma(q,\omega)=\frac{i8\pi e^2E_F\omega}{h^2q^2v_F^2}\left(\frac{1}{\sqrt{1-\frac{q^2v_F^2}{\omega^2}}}-1\right)
\label{conductivity}.
\end{equation}
Note that $\sigma(q,\omega)$ diverges as we approach the line ${\omega=qv_F}$ signaling the breakdown of linear response theory. This is also the reason why plasmon dispersion can not cut this line (see figure 1(a)). 

Note also that oscillating current will induce vector potential that will act back on the current. It is straight forward to solve Maxwell equations for the current oscillating in the plane of graphene ${j_x({\bf r},t)=j_0\sin{(\omega t-qx)}}$, and show that it will induce a vector potential: $A_x^{ind}({\bf r},t)=A_0^{ind}\sin{(\omega t - qx)}$, of the amplitude: 
\begin{equation}
A_0^{ind}=\frac{-qj_0}{\omega^2 2\varepsilon_0\varepsilon_r}, 
\label{Aind}
\end{equation}
where $\varepsilon_r=(\varepsilon_{r_1}+\varepsilon_{r_2})/2$ is the average dielectric constant of materials surrounding graphene from atop and below \cite{Jablan2009,Jablan2015}. If we then introduce some external potential $A_x^{ext}({\bf r},t)=A_0^{ext}\sin{(\omega t - qx)}$, the current will respond not only to $A^{ext}$ but to the total potential $A=A^{ext}+A^{ind}=A_0\sin{(\omega t - qx)}$ i.e. $j_x=i\omega\sigma(q,\omega)A$. The amplitude of this self-consistent potential is then: 
\begin{equation}
A_0=\frac{A_0^{ext}}{1+\frac{iq\sigma(q,\omega)}{\omega 2\varepsilon_0\varepsilon_r}}. 
\label{selfconsistent}
\end{equation}
One can see that it is possible to have self-sustained oscillations of the electron gas (plasmon-polaritons) even in the absence of the external field if: $1+\frac{iq\sigma(q,\omega)}{\omega 2\varepsilon_0\varepsilon_r}=0$, with the corresponding plasmon dispersion:
\begin{equation}
\omega(q)=qv_F\frac{ql+1}{\sqrt{ql(ql+2)}},
\label{dispersion}
\end{equation}
where we have introduced the length $l=\frac{\varepsilon_0\varepsilon_r h^2v_F}{4\pi e^2p_F}$.

In the local case ($q=0$) conductivity (\ref{conductivity}) reduces to $\sigma(\omega)=\frac{i}{\omega}\frac{e^2E_F}{\pi\hbar^2}$ in which case it is also easy to include losses (due to impurity of phonon scattering) via the phenomenological damping rate $\gamma$ as: $\sigma(\omega)=\frac{i}{\omega+i\gamma}\frac{e^2E_F}{\pi\hbar^2}$ \cite{Jablan2009}. It is usual to introduce the DC mobility $\mu$ via the following relation: $\sigma(0)=ne\mu$, so we can express damping rate as: $\gamma=\frac{ev_F^2}{\mu E_F}$.

\section{Landau-Zener model}
\label{LZ}

Let us focus on the state $\Psi_m({\bf r},t)=\sum_n c_n(u)\Psi_n^{QC}({\bf r},t)$, where $\Psi_n^{QC}=b_n\Psi_n^{qc}$ are our generalized quasiclassical states ($\Psi_n^{qc}$ multiplied by $b_n(u)$ also satisfies the quasiclassical condition). Now $\Psi_n^{QC}$ are asymptotically exact solutions as long as we are far away from the transition point $E_n^c(u_0)=E_{-n}^c(u_0)$, which is generally complex \cite{LLQM}. We can then connect these asymptotic states by going into complex $u$ plane, always staying far away from the transition point $u_0$ so that the quasiclassicality condition is always satisfied. This way $E_n^c$ from Eq. (\ref{energy}) simply changes the branch of the square root i.e. turns into $E_{-n}^c$, and similarly for other quantities. One can show that: ${|c_{-m}(u=\infty)|^2=\exp\left({\frac{1}{\hbar\omega} \textrm{im} \int_{C} \Delta E_m^c du}\right)=K}$, where the integration contour $C$ goes around the transition point $u_0$ in the upper half plane for $m=-1$, and around $u_0^*$ in the lower half plane for $m=1$ \cite{LLQM}. For convenience we write explicitly the energy gap:
\begin{equation}
\begin{split}
&\Delta E_m^c=E_m^c-E_{-m}^c\\
&=\frac{2mv_F}{1-\frac{q^2v_F^2}{\omega^2}}\sqrt{\left(p_m-\frac{q}{\omega}E_m-eA\right)^2+p_y^2\left(1-\frac{q^2v_F^2}{\omega^2}\right)},
\end{split}
\label{gap}
\end{equation}
where we have used the fact that: $p_{-m}-\frac{q}{\omega}E_{-m}=p_m-\frac{q}{\omega}E_m$.
Generally $u_0$ is complex, except in the case $p_y=0$ when we can have real $u_0$ and a perfect transition $K=1$. This is completely analogous to the famous Klein tunneling in graphene where electrons can simply pass through the potential barrier by using the available negative energy states \cite{Katsnelson2006}. 
Since $K$ is a probability of transition into a different band during a single passage, then $1-K$ is the probability that electron remains in the original band. As our field oscillates periodically in $u$, we also need to consider transition probability for a double passage: $w=K(1-K)+{(1-K)K}=2K(1-K)$ \cite{LLQM}. Finally, for very slow oscillations we can approximately say that transition happens at a real point $u$ where the gap $\Delta E_m^c(u)$ has a minimum, since then the tunneling probability is largest. This generally happens at two points $\zeta<\xi$ during a single period so we can write: $|c_{-m}(u)|^2\approx K\Theta(u-\zeta)\Theta(\xi-u)+2K(1-K)\Theta(u-\xi)$, where $\Theta(u)$ is a step function. Of course, to truncate dynamics to a single period only makes sense if $2K(1-K)\ll1$, which is the only regime we will explore in this paper. Finally since $\frac{d\Theta(u)}{du}=\delta(u)$ is a delta function, we can write: $\frac{d|c_{-m}(u)|^2}{du}=K\delta(u-\zeta)-K\delta(u-\xi)+2K(1-K)\delta(u-\xi)$. When calculating dissipated power, first two parts cancel and the only term that contributes is: 
\begin{equation}
\frac{d|c_{-m}(u)|^2}{du}=2K(1-K)\delta(u-\xi).
\end{equation}

\end{document}